\documentclass[runningheads]{svmult}

\usepackage{makeidx}   
\usepackage{graphicx}  
\usepackage{subeqnar}  
\usepackage{multicol}  
\usepackage{physprbb}  
\makeindex             

\begin{document}
\title*{Lattice versus Lennard-Jones models with a net particle flow}
\toctitle{Lattice versus Lennard-Jones models with a net particle flow}
%
%
\titlerunning{Lattice versus Lennard-Jones models}
%
\author{M. D\'iez-Minguito, P. L. Garrido, and J. Marro}
\authorrunning{M. D\'iez-Minguito et al.}
%
%
\institute{Institute `Carlos I' for Theoretical and Computational Physics, and\\
Departamento de Electromagnetismo y F\'{\i}sica de la Materia,\\
Universidad de Granada, E-18071 - Granada, Spain}

\maketitle              

\begin{abstract}
We present and study lattice and off-lattice microscopic models in which
particles interact via a local anisotropic rule. The rule induces preferential hopping along one direction, so
that a net current sets in if allowed by boundary conditions. This may be
viewed as an oversimplification of the situation concerning certain traffic
and flow problems. The emphasis in our study is on the influence
of dynamic details on the resulting (non-equilibrium) steady state. In
particular, we shall discuss on the similarities and differences between a
lattice model and its continuous counterpart, namely, a Lennard--Jones
analogue in which the particles' coordinates vary continuously. Our study,
which involves a large series of computer simulations, in particular reveals
that spatial discretization will often modify the resulting morphological
properties and even induce a different phase diagram and criticality.
\end{abstract}

\section{Introduction}
Many systems out of equilibrium \cite{haken,cross} exhibit spatial striped patterns on macroscopic scales. These are often caused by transport of matter or charge induced by a drive which leads to heterogeneous ordering. Such phenomenology occurs in flowing fluids \cite{rheology2}, and during
phase separation in colloidal \cite{loewen3}, granular \cite%
{reis,sanchez}, and liquid--liquid \cite{liqliq} mixtures. Further
examples are wind ripples in sand \cite{dunes3}, trails by animals and pedestrians \cite{helbing}, and the anisotropies observed in high temperature superconductors \cite{cuprates0,cuprates1} and in
two--dimensional electron gases \cite{2deg1,mosfet}.

Studies of these situations, often described as nonequilibrium phase transitions, have generally focused on lattice systems \cite{Liggett,Privman,Zia,Marro,Odor}, i.e., models based on a discretization of space and in considering interacting particles that move according to simple local rules. Such simplicity sometimes allows for exact calculations and is easy to be implemented in a computer. Moreover, some powerful techniques have been developed to deal with these situations, including nonequilibrium statistical field theory. However, lattice models are perhaps a too crude oversimplification of fluid systems so that the robustness of such an approach merits a detailed study.


The present paper describes Monte Carlo (MC) simulations and field theoretical calculations that aim at illustrating how slight modifications of dynamics at the microscopic level may influence, even quantitatively, the resulting (nonequilibrium) steady state. We are also, in particular concerned with the influence of dynamics on criticality. With this objective, we take as a reference the \textit{driven lattice gas} (DLG), namely, a kinetic nonequilibrium Ising model with conserved dynamics. This system has become a prototype for anisotropic behavior, and it has been useful to model, for instance, ionic currents \cite{Marro} and traffic flows \cite{antal}. In fact, in certain aspects, this model is more realistic for traffic flows than the standard \textit{asymmetric simple exclusion process} \cite{Liggett,Privman}. Here we compare the transport and critical properties of the DLG with those for apparently close lattice and off--lattice models. There is some related previous work addressing the issue of how minor variations in the dynamics may induce dramatic morphological changes both in the early time kinetics and in the stationary state \cite{valles,rutenberg,manolo0}. However, these papers do not focus on transport nor on critical properties. We here in particular investigate the question of how the lattice itself may condition transport, structural and critical properties and, with this aim, we consider nearest--neighbor (NN) and next--nearest--neighbor (NNN) interactions. We also compare with a microscopically off--lattice representation of the \textit{driven lattice gas} in which the particles' spatial coordinates vary continuously. A principal conclusion is that spatial discretization may change significantly not only morphological and early--time kinetics properties, but also critical properties. This is in contrast with the concept of universality in equilibrium systems, where critical properties are independent of dynamic details.

\begin{figure}[b]
\begin{center}
\includegraphics[scale=0.245]{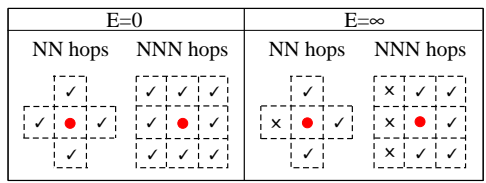}
\end{center}
\caption[]{Schematic comparison of the sites a particle (at the center, marked with a dot) may occupy (if the corresponding site is empty) for nearest--neighbor (NN) and next--nearest--neighbor (NNN) hops at equilibrium (left) and in the presence of an \textquotedblleft infinite\textquotedblright $ $ horizontal field (right). The particle--hole exchange between neighbors is either forbidden ($\times$) or allowed ($\surd$), depending on the field value.}
\label{fig1}
\end{figure}

\section{Driven Lattice Gases}
The \textit{driven lattice gas}, initially proposed by Katz, Lebowitz, and Spohn \cite{KLS}, is a nonequilibrium extension of the Ising model with conserved dynamics. The DLG consists of a \textit{d}-dimensional square lattice gas in which pair of particles interact via an attractive and short--range Ising--like Hamiltonian, 

\begin{equation}
H=-4\sum_{\langle j,k \rangle} \sigma_j \sigma_k \: .
\label{eq:ham}
\end{equation}
Here $\sigma_k=0(1)$ is the lattice occupation number at site $k$ for an empty (occupied) state and the sum runs over all the NN sites (the accessible sites are depicted in Fig.~\ref{fig1}). Dynamics is induced by the competion between a heat bath at temperature $T$ and an external driving field $E$ which favors particle hops along one of the principal lattice directions, say horizontally ($\hat{x}$), as if the particles were positively charged. Consequently, for periodic boundary conditions, a nontrivial nonequilibrium steady state is set in asymptotically. 
MC simulations by a biased \textit{Metropolis} rate reveal that, as in equilibrium, the DLG undergoes a second order phase transition. At high enough temperature, the system is in a disordered state while, below a critical point (at $T\leq T_{E}$) it orders displaying anisotropic phase segregation. That is, an anisotropic (striped for $d=2$) rich--particle phase then coexists with its gas. It is also found that the critical temperature $T_{E}$ monotonically increases with $E$. More specifically, for $d=2$, assuming a half filled square lattice in the large field limit (in order to maximize the nonequilibrium effect), one has a \textit{nonequilibrium} critical point at $T_{\infty}\simeq 1.4T_{0}$, where the equilibrium value is $T_{0}=2.269Jk_{B}^{-1}$. It was numerically shown that this belongs to a universality class other than the Onsager one, e.g., MC data indicates that the order parameter critical exponent is $\beta_{\mathrm{DLG}} \simeq 1/3$ \cite{Marro,achahbar} (instead of the Onsager value $1/8$).

\begin{figure}[b]
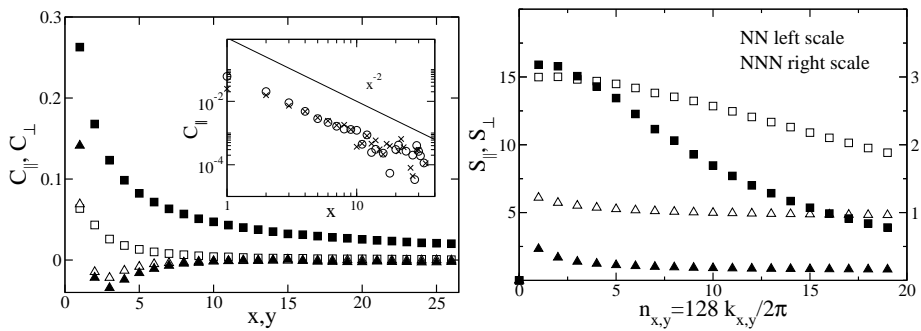

\begin{center}
\includegraphics[width=6cm]{fig2a.eps}
\includegraphics[width=6cm]{fig2b.eps}
\end{center}
\caption[]{Parallel (squares) and transverse (triangles) components of the two--point correlation function (left) and the structure factor (right) above criticality with NN (filled symbols) and NNN (empty symbols) interactions for a $128 \times 128$ half filled lattice. The inset shows the $x^{-2}$ power law decay in $C_{\parallel}$ for both discrete cases: DLG ($\circ$) and NDLG ($\times$).}
\label{fig2}
\end{figure}

Other key features concern the two--particle correlation function $C(x,y)$ and its Fourier transform $S(k_{x},k_{y})$, i.e., the structure factor. As depicted in the left graph of Fig.~\ref{fig2}, correlations are favored (inhibited) along (against) the field direction. In fact, the DLG shows a slow decay of the two--point correlations due to the spatial anisotropy associated with the dynamics \cite{pedro}. This long range behavior translates into a characteristic discontinuity singularity at the origin ($\lim_{k_{x}\rightarrow 0}S_{\parallel}\neq \lim_{k_{y}\rightarrow 0}S_{\perp}$) in the structure factor \cite{Zia}, which is confirmed in Fig.~\ref{fig2}.

How do all these features depend on the number of neighbor sites to which a particle can hop? Or in other words, how robust is the behavior when extending interactions and accessible sites to the NNN? 

Previous work has shown that extending hopping in the DLG to NNN leads to an inversion of triangular anisotropies during the formation of clusters \cite{rutenberg}, and also that dramatic changes occur in the steady state, including the fact that, contrary to the DLG with NN interactions, the critical temperature decreases with increasing $E$ \cite{manolo0}. However, other important features such as correlations and criticality seem to remain invariant. Analysis of the parallel ($C_{\parallel}$) and transverse ($C_{\perp}$) components reveals that correlations are quantitatively similar for the DLG and for the DLG with NNN interactions (henceforth NDLG) ---although somehow weaker for the latter case. Also persists a slow decay of correlations which yield to the discontinuity at the origin of $S(k_x,k_y)$. These facts are shown in Fig.~\ref{fig2}. 

On the other hand, recent MC simulations of the NDLG indicate that the order parameter critical exponent is $\beta_{\mathrm{NDLG}}\approx 1/3$ \cite{achahbar2}, as for the DLG. The \textit{anisotropic diffusive system} approach \cite{paco}, which is a Langevin--type (mesoscopic) description, predicts this critical behavior. In both cases, DLG and NDLG, the Langevin equations, as derived by coarse graining the master equation, lead to $\beta=1/3$. These two Langevin equations are identical, except for new entropic terms in the NDLG due to the presence of additional neighbors \cite{manolo2}. 

The fact that extending particle hops and interaction to the diagonal sites leaves invariant both correlations and criticality seems to indicate that the two systems, DLG and NDLG, belong to the same universality class. 

\section{A Driven Off-lattice Gas}
In order to deep further on this interesting issue, we studied to what extent the DLG behavior depends on the lattice itself. With this aim, we considered a driven system with continuous variation of the particles' spatial coordinates ---instead of the discrete variations in the DLG--- which follows as close as possible the DLG strategy. In particular, we analyzed an off--lattice, microscopically--continuum analog of the DLG with the symmetries and short--range interaction of this model.\\

\subsection{The Model}
Consider a \textit{fluid} consisting of $N$ interacting particles of mass $m$ confined
in a two--dimensional box of size $L\times L$ with periodic (toroidal) boundary
conditions. The particles interact via a truncated and shifted Lennard--Jones (LJ) pair potential \cite{Allen}:%
\begin{equation}
\phi(r)\equiv \left\{ 
\begin{array}{cl}
\phi_{LJ}(r)-\phi_{LJ}(r_{c}), & \mathit{if} \: r<r_{c} \\ 
0, & \mathit{if} \: r\geq r_{c},%
\end{array}%
\right.  
\label{pot}
\end{equation}%
where $\phi_{LJ}(r)=4\epsilon \left[ (\sigma /r)^{12}-(\sigma
/r)^{6}\right]$ is the LJ potential, $r$ is the interparticle distance, and $r_{c}$ is the 
\textit{cut-off} which we shall set at $r_{c}=2.5\sigma$. The parameters $\sigma$ and $\epsilon$ are, respectively, the characteristic length and energy. For simulations, all the quantities were reduced according to $\epsilon$ and $\sigma$, and $k_{B}$ and $m$ are set to unity. 

\begin{figure}[b]
\begin{center}
\includegraphics[width=9cm]{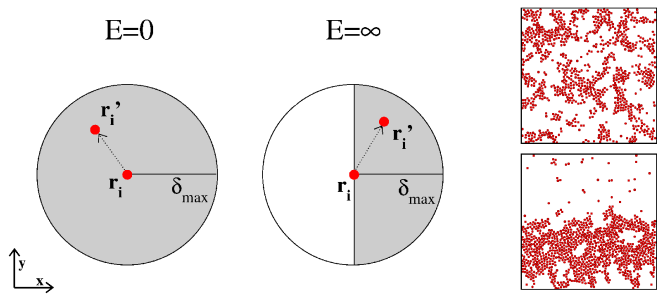}
\end{center}
\caption[]{Schematic representation of the accessible (shaded) region for a particle trial move at equilibrium (left) and out-of-equilibrium (right), assuming the field points along the horizontal direction ($\hat x$). The right hand side shows typical steady state configurations above (upper snapshot) and below (lower snapshot) criticality in the large field limit.}
\label{fig3}
\end{figure}

The uniform (in space and time) external driving field $E$ is implemented by assuming a preferential hopping in the horizontal direction. This favors particle jumps along the field, as it the particles were positively charged; see dynamic details in Fig.~\ref{fig3}. As in the lattice counterpart, we consider the large field limit $E\rightarrow \infty$. This is the most interesting case because, as the strength of the field is increased, one eventually reaches saturation, i.e., particles cannot jump against the field. This situation may be formalized by defining the transition probability per unit time (\textit{rate}) as
\begin{equation}
\omega(\eta \rightarrow \eta'; E, T)=\frac{1}{2}\left[   1+\tanh(E \cdot \delta) \right]   \cdot \min \left\lbrace 1,exp(-\Delta \Phi/T)\right\rbrace .
\label{rate}
\end{equation}
Here, any configuration is specified by $\eta\equiv\left\{ \mathbf{r}_{1},\cdots,\mathbf{r}_{N}\right\}$, where $\mathbf{r}_i$ is the position of the particle $i$, that can move anywhere in the torus, $\Phi(\eta)=\sum_{i<j}\phi(|\mathbf{r}_i-\mathbf{r}_j|)$ stands for the energy of $\eta$, and $\delta=(x_{i}^{\prime }-x_{i})$ is the displacement corresponding to a single MC trial move along the field direction, which generates an increment of energy $\Delta \Phi =\Phi(\eta^{\prime})-\Phi(\eta)$. The biased hopping which enters in the first term of Eq.~(\ref{rate}) makes the \textit{rate} asymmetric under $\eta \leftrightarrow \eta^{\prime}$. Consequently, Eq.~(\ref{rate}), in the presence of toroidal boundary conditions, violates detailed balance. This condition is only recovered in the absence of the driving field. In this limit the \textit{rate} reduces to the Metropolis one, and the system corresponds to the familiar \textit{truncated and shifted two--dimensional LJ fluid} \cite{smit,Allen}. Note that each trial move concerning any particle will satisfy that $0<|\mathbf{r}_{i}^{\prime}-\mathbf{r}_{i}|<\delta_{max},$ where $\delta_{max}$ is the maximum displacement in the radial direction (fixed at $\delta_{max}=0.5$ in our simulations). 

MC simulations using the \textit{rate} defined in Eq.~(\ref{rate}) show highly anisotropic states (see Fig.~\ref{fig3}) below a critical point which is determined by the pair of values $(\rho_{\infty},T_{\infty})$. A linear interface forms between a high density phase and its vapor: a single strip with high density extending horizontally along $\hat{x}$ throughout the system separates from a lower density phase (vapor). The local structure of the anisotropic condensate changes from a strictly hexagonal packing of particles at low temperature (below $T=0.10$), to a polycrystalline--like structure with groups of defects and vacancies which show a varied morphology (e.g., at $T=0.12$), to a fluid--like structure (e.g., at $T=0.30,$) and, finally, to a disordered state as the temperature is increased further. This phenomenology makes our model useful for interpreting structural and phase properties of nonequilibrium fluids, in contrast with lattice models, which are unsuitable for this purpose. Skipping the microscopic structural details, the stationary striped state is similar to the one in lattice models, however.

\begin{figure}[b]
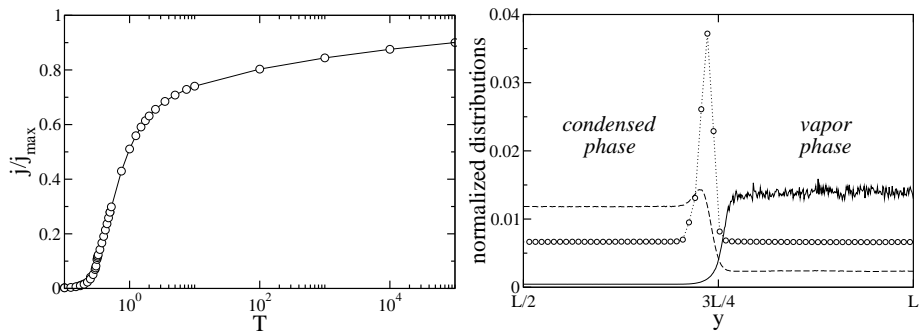

\begin{center}
\includegraphics[width=6cm]{fig4a.eps}
\includegraphics[width=6cm]{fig4b.eps}
\end{center}
\caption[]{Left graph: Temperature dependence of the net current for the driven LJ fluid. Right graph: Transverse--to--the--field current profiles below criticality. The shaded (full) line corresponds to the current (velocity) profile of the off--lattice model. For comparison we also show the current profile of the DLG with NN interactions (circle--dotted line). Since each distribution is symmetric with respect to the system center of mass (located here at $L/2$) we only show their right half parts.}
\label{fig4}
\end{figure}

\subsection{Transport Properties}
Regarding the comparison between off--lattice and lattice transport properties, the left graph in Fig.~\ref{fig4} shows the net current $j$ as a function of temperature. Saturation is only reached at $j_{max}=4\delta_{max}/3\pi$ when $T\rightarrow \infty$. The current approaches its maximal value logarithmically, i.e., slower than the exponential behavior predicted by the Arrhenius law. The sudden rising of the current as $T$ is increased can be interpreted as a transition from a poor--conductor (low--temperature) phase to a rich--conductor (high--temperature) phase, which is reminiscent of ionic currents \cite{Marro}. This behavior of the current also occurs in the DLG. Revealing the persistence of correlations, the current is nonzero for any low $T,$ though very small in the solid--like phase. From the temperature dependence of $j$ one may estimate the transitions points between the different phases, in particular, as the condensed strip changes from solid to liquid ($T\approx 0.15$) and finally changes to a fully disordered state ($T\approx 0.31$).

The current is highly sensitive to the anisotropy. The most relevant information is carried by the transverse--to--the--field current profile $j_{\perp}$, which shows the differences between the two coexisting phases (right graph in Fig.~\ref{fig4}). Above criticality, where the system is homogeneous, the current profile is flat on the average. Otherwise, the condensed phase shows up a higher current (lower mean velocity) than its mirror phase, which shows up a lower current (higher mean velocity). Both the transversal current and velocity profiles are shown in Fig.~\ref{fig4}. The current and the density vary in a strongly correlated manner: the high current phase corresponds to the condensed (high density) phase, whereas the low current phase corresponds to the vapor (low density) phase. This is expectable due to the fact that there are many carriers in the condensed phase which allow for higher current than in the vapor phase. However, the mobility of the carriers is much larger in the vapor phase. The maximal current occurs in the interface, where there is still a considerable amount of carriers but they are less bounded than in the particles well inside the \textit{bulk} and, therefore, the field drives easily those particles. This enhanced current effect along the interface is more prominent in the lattice models (notice the large peak in the current profile in Fig.~\ref{fig4}). Moreover in both lattice cases, DLG and NDLG, there is no difference between the current displayed by the coexisting phases because of the particle--hole symmetry. Such a symmetry is derived from the Ising--like Hamiltonian in Eq.~(\ref{eq:ham}) and it is absent in the off--lattice model.



\begin{figure}[b]
\begin{center}
\includegraphics[width=6cm]{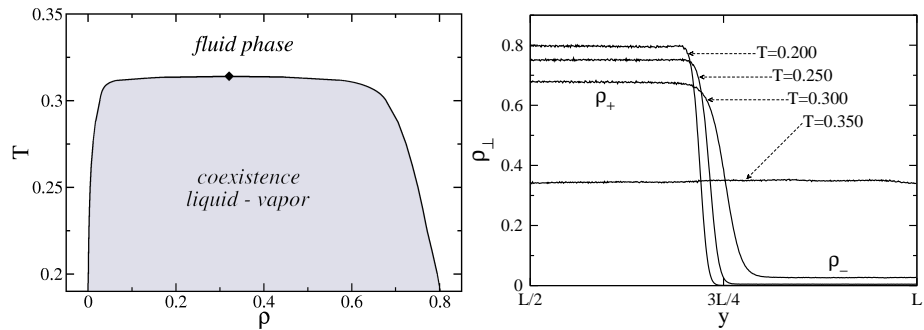}
\includegraphics[width=6cm]{fig5b.eps}
\end{center}
\caption[]{The temperature--density phase diagram (left graph)was obtained from the transversal density profile (right graph) for $N=7000,$ $\protect\rho =0.35,$ and different temperatures. The coexistence curve separates the liquid--vapor region (shaded area) and the liquid phase (unshaded area). The diamond represents the critical point, which has been estimated using the scaling law and the rectilinear diameter law (as defined in the main text).}
\label{fig5}
\end{figure}

\subsection{Critical Properties}
A main issue is the (nonequilibrium) liquid--vapor coexistence curve and the associated critical behavior. The coexistence curve may be determined from the density profile transverse to the field $\rho_{\perp}$. This is illustrated in Fig.~\ref{fig5}. At high enough temperature above the critical temperature the local density is roughly constant around the mean system density ($\rho=0.35$ in Fig.~\ref{fig5}). As $T$ is lowered, the profile accurately describes the striped phase of density $\rho _{+}$ which coexists with its vapor of density $\rho _{-}$ ($\rho _{-}\leq\rho _{+}$). The
interface becomes thinner and less rough, and $\rho _{+}$ increases while $\rho_{-}$ decreases, as $T$ is decreased. As an order parameter for the second order phase transition one may use the difference between the coexisting densities $\rho _{+}-\rho _{-}$. The result of plotting $\rho _{+}$ and $\rho _{-}$ at each temperature is shown in Fig.~\ref{fig5}. The same behavior is obtained from the transversal current profiles (Fig.~\ref{fig4}). It is worth noticing that the estimate of the coexisting densities $\rho_{\pm}$ is favored by the existence of a linear interface, which is simpler here than in equilibrium. This is remarkable because we can therefore get closer to the critical point than in equilibrium. 

Lacking a \textit{thermodynamic} theory for \textquotedblleft phase transitions\textquotedblright $ $ in non--equilibrium liquids, other approaches have to be considered in order to estimate the critical parameters. Consider to the rectilinear diameter law $(\rho _{+}+\rho _{-})/2=\rho _{\infty }+b_{0}(T_{\infty }-T)$ which is a empirical fit extensively used for fluids in equilibrium. This, in principle, has no justification out of equilibrium. However, we found that our MC data nicely fit the diameters equation. We use this fact together with a universal scaling law $\rho _{+}-\rho _{-}=a_{0}(T_{\infty }-T)^{\beta }$ to accurately estimate the critical parameters. The simulation data in Fig.~\ref{fig5} thus yields $\rho_{\infty}=0.321(5)$, $T_{\infty}=0.314(1)$, and $\beta=0.10(8)$, where the estimated errors in the last digit are shown in parentheses. These values are confirmed by the familiar log--log plots. Compared to the
equilibrium case \cite{smit}, one has that $T_{0}/T_{\infty }\approx 1.46$. This confirms the intuitive observation above that the field acts in this system favoring disorder. On the other hand, our estimate for the order--parameter critical exponent is fully consistent
with both the extremely flat coexistence curve which characterizes the
equilibrium two--dimensional LJ fluids and the equilibrium Ising value, $%
\beta_{\mathrm{Ising}} =1/8$ (non--mean--field value). Although the error bar is large, one may discard with confidence the DLG value $%
\beta_{\mathrm{DLG}} \approx 1/3$ as well as the mean field value. This result is striking because our model \textit{seems} to have the symmetries and short--range interactions of the DLG. Further understanding for this difference will perhaps come from the statistical field theory.




\section{Final Comments}
In summary, we reported MC simulations and field theoretical calculations to study the effect of discretization in \textit{driven diffusive systems} In particular, we studied structural, transport, and critical properties on the \textit{driven lattice gas} and related non--equilibrium lattice and off--lattice models. Interestingly, the present \textit{Lennard--Jones} model in which particles are subject to a constant driving field is a computationally convenient prototypical model for anisotropic behavior, and reduces to the familiar LJ case for zero field. Otherwise, it exhibits some arresting behavior, including currents and striped patterns, as many systems in nature. We have shown that the additional spatial freedom that our fluid model possesses, compared with its lattice counterpart, is likely to matter more than suggested by some naive intuition. In fact, it is surprising that its critical behavior is consistent with the one for the Ising equilibrium model but not with the one for the \textit{driven lattice gas}. The main reason for this disagreement might be the particle--hole symmetry violation in the driven \textit{Lennard--Jones} fluid. However, to determine exactly this statement will require further study. It also seems to be implied that neither the current nor the inherent anisotropy are the most relevant feature (at least regarding criticality) in these driven systems. Indeed, the question of what are the most relevant ingredients and symmetries which determine unambiguously the universal properties in driven diffusive systems is still open. In any case, the above important difference between the lattice and the off--lattice cases results most interesting as an unquestionable nonequilibrium effect; as it is well known, such microscopic detail is irrelevant to universality concerning equilibrium critical phenomena.\\


We acknowledge very useful discussions with F. de los
Santos and M. A. Mu\~{n}oz, and financial support from MEyC and FEDER (project FIS2005-00791).

%

\end{document}